\documentclass[11pt]{article}

\def\M{{\cal M}}

\def\A{{\cal A}}
\def\F{{\cal F}}

\def\CP2{{{\bf CP}^2}}
\def\Dirac{{D\kern -6pt \raise 0pt \hbox{/}\,}} % Dirac operator
\def\Dots{\cdot\cdot}

\usepackage[matrix,arrow]{xy}

\usepackage{amsmath, amssymb, amsbsy, amsthm}

\usepackage{graphicx}

\input{xy}

\newcommand{\Tr}{\,{\rm Tr}\,}
\newcommand{\tr}{\,{\rm tr}\,}

\usepackage{bm}%bold math package (for bold greek)
\usepackage[vcentermath,enableskew]{youngtab}
\usepackage[english]{babel}
\usepackage{epsf}
\usepackage[latin1]{inputenc}
\usepackage{times}
\usepackage[T1]{fontenc}

\title{Equivariant Dimensional Reduction and Quiver Gauge
  Theories}
\author%[Brian Dolan] % (optional, use only with lots of authors)
{Brian~P.~Dolan$^1$ and Richard~J.~Szabo$^2$ \\[2mm]
$^1$ {\small\it National University of Ireland Maynooth, Ireland} \\
$^2$ {\small\it Heriot-Watt University, Edinburgh, Scotland }}

\begin{document}

\maketitle

\begin{abstract}
We review recent applications of equivariant dimensional reduction
techniques to the
construction of Yang-Mills-Higgs-Dirac theories with dynamical mass
generation and exactly massless chiral fermions. (Based on invited talk given by the first author at the
  2nd School on ``Quantum Gravity and Quantum Geometry''
session of the 9th Hellenic School on Elementary Particle Physics and
Gravity, Corfu, Greece, September 13--20 2009. To be published
in {\sl General
Relativity and Gravitation}.)
\end{abstract}

\section{A brief history of dimensional reduction}

The idea that the observed fundamental forces in 4-dimensions
can be understood in terms of the dynamics of a simpler higher dimensional
theory is now nearly 90 years old~\cite{KK}.
Starting from a 5-dimensional theory on a manifold
$\M_5={\cal M}_4\times S^1$, where  ${\cal M}_4$ is a curved
4-dimensional space-time and the fifth dimension is
a perfect circle with radius $r$, and taking the 5-dimensional line element to be
 {{$(0\le y<2\pi)$}}:
\[
d s_{(5)}^2=ds_{(4)}^2 + \big(r dy +A(x)\big)^2,\]
where $A(x)=A_\mu(x)dx^\mu$ is a 4-dimensional vector potential,
the 5-dimensional Einstein action reduces to 
\[{ \frac{1}{2\pi r}\int_{\M_5} \sqrt{-g_{(5)}}\,{\cal R}_{(5)}\,d^4x\, dy
=\int_{{\cal M}_4}\sqrt{-g_{(4)}}\Big({\cal R}_{(4)}-\frac{1}{4}{F^2}\Big)d^4 x,}
\]
where $F=dA$ is a $U(1)$ field strength in 4-dimensions
and $F^2=F_{\mu\nu}F^{\mu\nu}$.

If we now introduce extra matter, e.g. a scalar field $\Phi$,
and perform a harmonic expansion on $S^1$,
\[ 
\Phi(x,y)=\sum_{n=-\infty}^\infty \phi_n(x) e^{\frac{i n y}{r}},
\]
then the 5-dimensional kinetic term for $\Phi$ gives rise
to an infinite tower of massive fields in 
${\cal M}_4$, $\phi_n(x)$, with masses $m_n= \frac{n}{r}$.

A non-abelian generalisation of the Kaluza-Klein idea uses
a $d$-dimensional manifold $\M_d={\cal M}_4\times S/R$, with
$R \subset S$ compact Lie groups. The co-set space $S/R$
has isometry group
$S$ and holonomy group $R$. Performing the integral $\int_{S/R} d\mu$
over the internal space, with $d\mu$ the $S$-invariant measure on $S/R$, leads to 
Yang-Mills 
gauge theory in 4-dimensions with gauge group $S$;
{e.g. $S^2\simeq SU(2)/U(1)$, with
$SU(2)$ isometry and $U(1)$ holonomy}, gives 4-dimensional
Einstein-Yang-Mills theory
with gauge group $SU(2)$, see {e.g.}~\cite{KKReview}.

Alternatively, one can start from $d$-dimensional Yang-Mills theory
on ${\cal M}_4 \times S/R$ with gauge group $G$.
Forg\'acs and Manton~\cite{FM} showed that interesting symmetry breaking
effects can occur if $R\subset G$ and one chooses a specific embedding 
$R\hookrightarrow G$.  Integrating over ${S/R}$ then gives
a Yang-Mills-Higgs system on $\M_4$, with a
gauge group $K$ which is the {centraliser} of $R$ in $G$, {i.e.}
$K\subset G$ with $[R,K]=0$ (see also~\cite{ThCh}).
Upon dimensional reduction the 
internal components of the $d$-dimensional gauge field $\A$
play the r\^ ole of Higgs fields in $4$-dimensions and a
Higgs potential is generated from the $d$-dimensional Yang-Mills
action:
\[ \A(x,y)\quad\longrightarrow\quad  
\left\{\begin{array}{lc} A_\mu(x) & {\hbox{(4-dimensional gauge fields)}}\\ 
\Phi_a(x) & {\hbox{(4-dimensional Higgs fields)}} \\ \end{array}\right.\]
{(here $x^\mu$ are co-ordinates on ${\cal M}_4$, $y^a$ co-ordinates on $S/R$)}.
The full $d$-dimensional Yang-Mills action, with field strength ${\cal F}$,
reduces as
\begin{eqnarray} 
-\frac 1 4 \int_{\M_d}\sqrt{-g_{(d)}}\,\Tr(\F^2)d^4 x\, d^{d-4} y =
{\rm vol}(S/R)\int_{\M_4}\sqrt{-g_{(4)}}\,\tr\Big(-\frac 1 4 F^2 + \bigl(D\Phi\bigr)^\dagger D\Phi - V(\Phi)\Big)d^4x,\nonumber  \end{eqnarray}
where $\Tr$ denotes trace over the $d$-dimensional gauge group $G$ 
and $\tr$ is over the $4$-dimensional gauge group $K$. Furthermore
the Higgs potential can break $K$ dynamically.
In particular if $S\subset G$, then $V(\Phi)$ breaks $K$ spontaneously to $K'$,
the centraliser of $S$ in $G$, $[S,K']=0$.

Consider again the simplest case $S^2\simeq SU(2)/U(1)$, where $S\cong SU(2)$
and $R\cong U(1)$.
For example if  $G=SU(3)$ then indeed $S\subset G$ and
in the first step $R \hookrightarrow G$:
$U(1)\hookrightarrow SU(3)$ breaking $SU(3)$ to $K=SU(2)\times U(1)$.
Upon reduction the $4$-dimensional Higgs doublet, 
$\Phi_a$, $a=1,2$, dynamically breaks $SU(2)\times U(1) \rightarrow 
K'\cong
U(1)$, which is the centraliser of $S=SU(2)$ in $G=SU(3)$.
Going beyond $SU(2)$ symmetry on the co-set space, a
harmonic expansion of, for example, a scalar field $\Phi$
on $S^2\simeq SU(2)/U(1)$,
\[\Phi(x,y)=\sum_{l=0}^\infty \sum_{m=-l}^l \phi_{l;m}(x)Y_l^m(y),\] 
generates a tower of higher modes, 
$\phi_{l;m}(x)$, which have masses $M^2_l=\frac{l(l+1)} {r^2}$
in $4$-dimensions.

Much of the steam was taken out of the co-set space
dimensional reduction programme with Witten's proof that
spinors on ${\cal M}_4 \times S/R$ {\it cannot} give a chiral theory
on ${\cal M}_4$~\cite{ShelterIsland}.

Reviews of co-set space dimensional reduction are given in
\cite{George} and \cite{Yuri}.

\section{Equivariant dimensional reduction}

\subsection{General construction} 

Equivariant dimensional reduction is a
systematic procedure for including internal fluxes on 
$S/R$ (instantons and/or monopoles of $R$-fields) which are \lq
symmetric' {(equivariant)} under $S$~\cite{ALCP,LPS}.
It relies on the fact that, with suitable restrictions on $S$ and $R$, 
there is a one-to-one correspondence
between $S$-equivariant complex vector bundles over ${\cal M}_d$
\[B\longrightarrow  {\cal M}_d = {\cal M}_4 \times S/R, \]
and $R$-equivariant bundles over ${\cal M}_4$,
\[E\longrightarrow  {\cal M}_4,\]
where $S$ acts on the space ${\cal M}_d$ via the trivial action on
${\cal M}_4$ and by the standard left translation action on $S/R$
(we shall restrict ourselves to the case where $S$ and $R$ are compact
and the embedding $R\hookrightarrow S$ is maximal).
If $B$ and $E$ are ${\bf C}^k$ vector bundles there is a commutative
diagram of bundle maps
\[
\xymatrix{
{\bf C}^k \ \ar[r]_R & \ E \ \ar[d] \ar[r]^{\rm induce} &\  B \ar[d] \  &
\ar[l]^S \ {\bf C}^k \\
& \M_4 \ & \ar[l]^{\rm restrict} \ \M_d &
}
\]
where the induction map is defined by
\[h\in R, \quad (g,e)\in S\times E, 
\qquad h\cdot(g,e)=(gh^{-1},he)\mapsto B.\]

In general the reduction gives rise to quiver gauge theories on $\M_4$.
Including spinor fields, coupling to background equivariant
fluxes, can give rise to chiral theories on $\M_4$.
One expects zero modes of the Dirac operator on $S/R$  
to manifest themselves as 
massless chiral fermions in $\M_4$ but, as we shall see,
Yukawa couplings are induced and the dimensional reduction
can give masses to some zero modes~\cite{DR1,DR2}.

\subsection{A simple example: Complex projective line}

Consider once again the simplest non-trivial example with $S\cong SU(2)$ and
$R\cong U(1)$, giving a 2-dimensional sphere
$S^2\simeq SU(2)/U(1)$ (or projective line ${\bf CP}^1$), and with $G\cong U(k)$. 
Choosing an embedding $S\hookrightarrow G$ gives a decomposition
$U(k)\rightarrow \prod_{i=0}^m U(k_i)$, where $k=\sum_{i=0}^m k_i$,
associated with the $m+1$-dimensional irreducible representation of $SU(2)$.
Let ${\bf g}\in G$,
${\bf v}\in {\bf C}^k$ and ${\bf v}_i\in {\bf C}^{k_i}$.  
Then, as a $k\times k$ matrix, ${\bf g}$ decomposes as
\[
{\bf g} ={\scriptstyle m+1} \left\{ \vline height 15pt depth 20pt width 0pt  \right.\overbrace
{\left(\begin{array}{cccc}
 {\bf g}_{k_0\times k_0} & {\bf g}_{k_0\times k_1} & \cdots & {\bf g}_{k_0\times k_m} \\
\vdots & \vdots & \ddots & \vdots\\
 {\bf g}_{k_m\times k_0} & {\bf g}_{k_m\times k_1} & \cdots & {\bf g}_{k_m\times k_m} \\
\end{array}\right)}^{m+1}\ ,\quad
 {\bf v}=\left(\begin{array}{c}
{\bf v}_0 \\ {\bf v}_1 \\ \vdots \\ {\bf v}_m \\
\end{array}\right),\]
where $SU(2)$ acts on ${\bf g}$ as a $(m+1)\times (m+1)$ block matrix.
Each subspace
${\bf v}_i$ transforms under $U(k_i)\subset U(k)$ and carries
a $U(1)$ charge $p_i=m-2i$, $-m\le p_i \le m$.

Introducing a complex co-ordinate $y$ on $S^2$ {(of radius $r$)},
\[ 
ds_{(2)}^2= r^2\beta \overline\beta,\qquad
\beta = \frac{2 d y}{1+y \overline y}, \]
we write the potential and field strength for a monopole of charge $p$
in these co-ordinates as
\[a_p=\frac{ip(y d\overline y  - \overline y d y)}{2(1+y\overline y)},\quad
f_p=\frac {ip} 4 \beta\wedge\overline\beta,\quad
\frac{1}{2\pi}\int_{S^2} f_p=p. %\quad {\hbox{charge of } L^p}.
\]
The $U(k)$ gauge potential, a Lie algebra valued 1-form $\A$ on $\M_d$, now splits
into $k_i\times k_j$ blocks
\[ \ \hskip -20pt {\A(x,y) = A(x) + a(y) + \Phi(x)\overline\beta(y) + \Phi^\dagger(x)\beta(y),}\]
where $A=\oplus_{i=0}^mA^i$, $a=\oplus_{i=0}^ma_{m-2i}$,
$A^i(x)$ is a $U(k_i)$ gauge connection on $\M_4$, and $\Phi(x)$ will acquire the
interpretation as a set of Higgs fields.
As a $(m+1)\times (m+1)$ block matrix
\[ \ \hskip -30pt \A(x,y) = \left(\begin{array}{ccccc}
A^0+a_m\,{\bf 1}_{k_0} & \phi_1\overline\beta & 0 & \cdots & 0 \\
 \phi_1^\dagger\beta & A^1+a_{m-2}\,{\bf 1}_{k_1} & \phi_2\overline\beta & \cdots & 0 \\
\vdots & \vdots &\vdots  & \ddots & \vdots \\
0 & 0 & 0 & \cdots & \phi_m\overline\beta \\
0 & 0 & 0 & \cdots & A^m+a_{-m}\,{\bf 1}_{k_m} \\ 
\end{array}\right),\]
where each $\phi_i$ is a $k_{i-1}\times k_i$ matrix
transforming under $U(k_{i-1})_L\times U(k_i)_R$.
As a $(m+1)\times (m+1)$ matrix the 
Higgs field is
\[ \Phi=\left(\begin{array}{ccccc}
0 & \phi_1 & 0 & \cdots & 0 \\
0 & 0 & \phi_2 & \cdots & 0 \\
\vdots & \vdots & \vdots & \ddots & \vdots\\
0 & 0 & 0 & \cdots & \phi_m \\
0 & 0 & 0 & \cdots & 0 \\
\end{array}\right).\]

Dimensional reduction generates a 4-dimensional Higgs potential,
\[\kern -40pt  V(\Phi)=\frac {g^2} 2 \tr_k \Bigg(
\frac 1 {4 g^2 r^2}\left(\begin{array}{cccc} m{\bf 1}_{k_0} & 0 & \cdots & 0 \\
0 & (m-2){\bf 1}_{k_1} & \cdots & 0 \\
\vdots & \vdots & \ddots & \vdots \\
0 & 0 & 0 & -m{\bf 1}_{k_m} \\ \end{array}\right) -[\Phi,\Phi^\dagger]\Bigg)^2, \]
where $g$ is the 6-dimensional gauge coupling. The minimisation of the Higgs potential gives a
vacuum structure that depends on the monopole charges $p_i=m-2i$.

\subsubsection{Example: $\boldsymbol{SU(3)\rightarrow SU(2)\times U(1)\rightarrow U(1)}$}
\label{threetwoone}

As a concrete example, consider the case with $G\cong SU(3)$ and
$m=1$ (fundamental of $SU(2)$), so that $k=3$ and $k_0=2$, $k_1=1$.
In this case there is one unit charge monopole and one anti-monopole sector
in the internal space which give a symmetry breaking pattern
\[ SU(3)\ \xrightarrow{\rm reduction} \ 
SU(2)\times U(1) \ \xrightarrow{\rm dynamics} \  U(1), \]
so $K\cong SU(2)\times U(1)$ is broken dynamically
to $U(1)$ (for details, see \cite{DR1}).

There is only one Higgs multiplet, $\phi$, which is a 2-component vector,
and the minimum of $V(\phi)$ is at $\phi_0=\begin{pmatrix} 0 \\ \frac
  1 {2gr} \end{pmatrix}$
in a suitable gauge.
Perturbing around this vacuum gives
$\phi=\begin{pmatrix} 0 \\ \frac{1}{2gr} +h \end{pmatrix}$,
with $h$ real,
and the Higgs mass works out to be $m_h=\frac{1}{r}$.

The three gauge boson masses are $m_{W^
\pm}=\frac 1 2 m_Z=\frac 1 {\sqrt{2}r}$
while the Weinberg angle evaluates to $\sin^2\theta_W=\frac 3 4$. 
Clearly this is not a phenomenologically viable model for electroweak
interactions, as the gauge boson masses
and the Weinberg angle are wrong, but it is nevertheless instructive.

\subsubsection{Example: $\boldsymbol{SU(3k')\rightarrow SU(k')}$}

As a second example take $G\cong SU(k)$.
Let $m=2$ (adjoint of $SU(2)$) and choose
$k_0=k_1=k_2=k'$, so that $k=3k'$. 
There are now three sectors in the internal space, one charge two monopole,
its anti-monopole, and a trivial sector.
The symmetry breaking scheme in this case
is
\[ SU(3k')\ \xrightarrow{\rm reduction} \ 
SU(k')^3\times U(1)^2 \ \xrightarrow{\rm dynamics} \  SU(k')_{diag}.\]
There are two Higgs multiplets, $\phi_1$ and $\phi_2$, both of which are
$k\times k$ matrices. The Higgs potential is
\[V(\Phi)=
g^2\tr_k\big((\phi_1{}^\dagger \phi_1)^2 
-\phi_1{}^\dagger \phi_1\phi_2{}^\dagger \phi_2 +(\phi_2{}^\dagger \phi_2)^2 \bigr)
 - \frac 1 {2r^2} \tr_k\bigl(\phi_1{}^\dagger \phi_1+ \phi_2{}^\dagger \phi_2\bigr),\]
and we expand $\phi_i$ around the vacuum as
\[\phi_i=\frac{\sqrt{i(3-i})}{2gr} {\bf 1}_k+h_i,\] 
with $h_i=h_i^\dagger, i=1,2$.

Diagonalising the Higgs mass matrix produces two distinct
eigenvalues  $m_h^2=\frac{3}{r^2}$, $\frac 1 {r^2}$.
There are $k^{\prime 2}-1$ gauge bosons with mass $m_W^2=\frac 1 {2r^2}$,
$k^{\prime 2}-1$ with $m_{W'}^2=\frac 3 {2r^2}$, while two $Z$-bosons acquire masses 
$m_Z^2=\frac 1 {4r^2}$ and $m_{Z'}^2=\frac 9 {4r^2}$.

\subsubsection{Quiver diagrams}

This construction generates quiver gauge theories on $\M_4$.
Writing the Lie algebra of $SU(2)$ in the form $[J_3,J_\pm]=\pm 2 J_\pm$,
the Higgs fields give rise to a 
chain of bundle maps $\Phi_i$:
\[
 0 \hskip 5pt  \longrightarrow \hskip 5 pt
{B_0}\hskip 10pt
\xrightarrow{\Phi_1}
\hskip 10pt {B_1} 
\hskip 10pt \xrightarrow{\Phi_2} \hskip 10pt \cdots  
\hskip 10pt \xrightarrow{\Phi_{m-1}} \hskip 10pt 
{B_{m-1}}  
\hskip 10pt \xrightarrow{\Phi_m} \hskip 10pt 
{B_m}  \hskip 5pt  \longrightarrow  \hskip 5pt  0. \]
\begin{center}
\epsfxsize 9cm
\hskip -10pt \epsffile{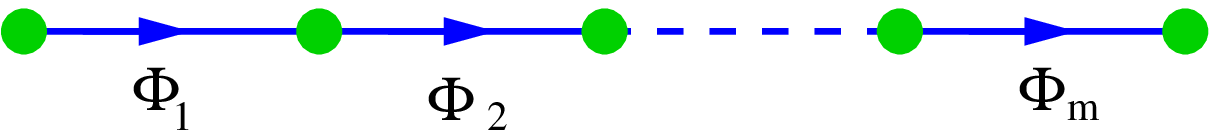}
\end{center}
The isometry group $SU(2)$ is rather special in that there is only one raising and
one lowering operator, so the quiver diagram is always a chain.
Higher rank isometry and holonomy groups generate more complicated quiver diagrams
in general.

\subsection{A more general example: Complex projective plane}

As a more general example consider
{${\bf CP}^2\simeq SU(3)/U(2)$} (for details see \cite{LPS}
and \cite{DR2}). Label the irreducible representations of $SU(3)$ by $\{l,\overline l\}$,
corresponding to the Young tableau
$$\underbrace{\young(\ \Dots\ ,\  \Dots \ )}_{\overline l}\kern -2.1pt
\raise 6.2pt\hbox{$\underbrace{\young(\ \Dots\ )}_l$}$$ Denote irreducible representations of $SU(2)\times U(1)$ by $(n,m)$, with $n=2I$ {(isospin)} and 
$m=3Y$ {(hypercharge)}.
Then under the embedding
$U(2)\hookrightarrow SU(3)$, the irreducible representations decompose as 
$\{l,\overline l\}\rightarrow \oplus(n,m):=W_{l,\overline l}$,
where $W_{l,\overline l}$ represents the set of all $SU(2)\times U(1)$ irreducible representations
in $\{l,\overline l\}$.
For example, $W_{1,0}$ has two elements: ${\bf 3}\rightarrow {\bf 2}_1 \oplus {\bf 1}_{-2}$.

The root diagram for $SU(3)$ is
\begin{center}
\epsfxsize 5cm\epsffile{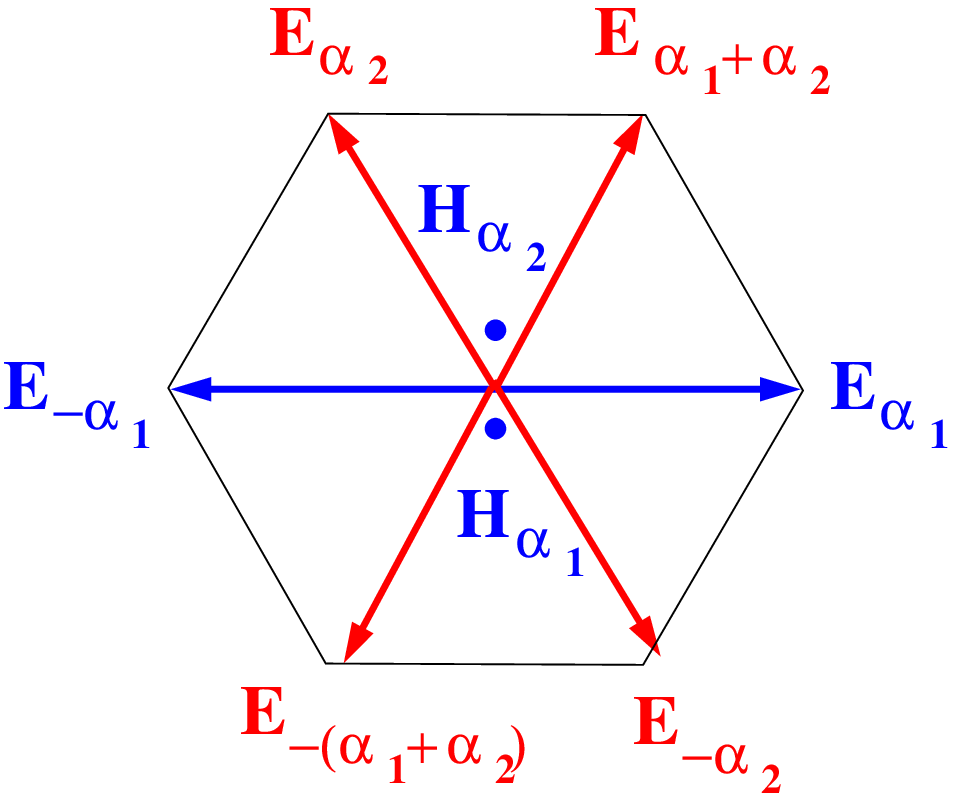}
\end{center}
For any given irreducible representation $\{l,\overline l\}$,
$E_{\alpha_2}$ and $E_{\alpha_1+\alpha_2}$ map between elements of
$W_{l,\overline l}$ with different isospin and can be decomposed into 
components that increase the isospin and components that decrease it:
\[E_{\alpha_2}=E^+_{\alpha_2}+E^-_{\alpha_2},\qquad
E_{\alpha_1+\alpha_2}=E^+_{\alpha_1+\alpha_2}+E^-_{\alpha_1+\alpha_2},\]
with
\[E^\pm_{\alpha_2}:(n,m)\longrightarrow (n\pm1,m+3),\ \quad
E^\pm_{\alpha_1+\alpha_2}:(n,m)\longrightarrow (n\pm1,m+3).\]

Choosing a basis of orthonormal 1-forms for $\CP2$
which is compatible with the complex structure,
$\beta^1$, $\beta^2$, $\overline \beta^1$, $\overline \beta^2$,
define the Lie-algebra valued 1-forms $\beta^\pm_{n,m}$,
together with their complex conjugates,
via the relations
\[  \beta^\pm: = \beta^1 E^\pm_{\alpha_1+\alpha_2}+\beta^2 E^\pm_{\alpha_2}=\sum_{(n,m)\in W_{l,\overline l}}\beta^\pm_{n,m}.
\]
There is then a Higgs field, $\phi_{n,m}^\pm$,
associated with each $\beta^\pm_{n,m}$.

\subsubsection{Example: Adjoint representation}

For example, the adjoint representation $l=\overline l=1$ of $SU(3)$
decomposes as 
\[W_{1,1}=\big\{\stackrel{ K}{(1,3)}\oplus \stackrel{{\overline K}}{(1,-3)}
\oplus \stackrel{{\bm\pi} }{(2,0)}\oplus \stackrel{{\bm{\eta}}}{(0,0)}\big\},\]
where the different $SU(2)\times U(1)$ representations are also indicated by their
usual particle physics notation.
Choosing the gauge group to be
\[G=U(k)\longrightarrow U(k_{1,3})\times U(k_{1,-3})\times U(k_{2,0})\times
U(k_{0,0}),\]
with $k=2k_{1,3}+2k_{1,-3}+3k_{2,0}+k_{0,0},$
there are four Higgs fields mapping between the  $SU(2)\times U(1)$ representations
and the quiver diagram assumes the form
\begin{center}
\epsfxsize 3cm\epsffile{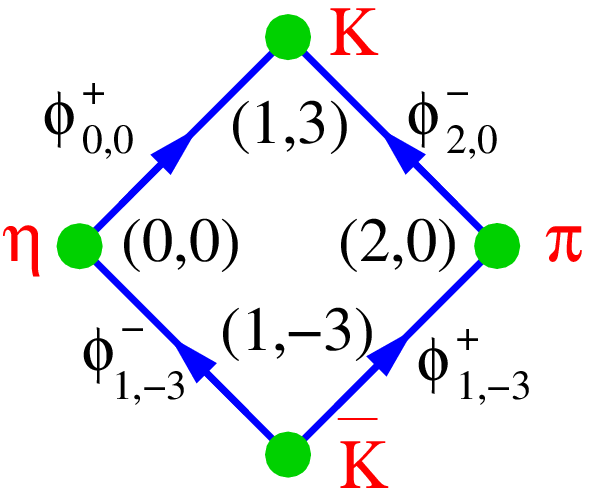}
\end{center}

For illustrative purposes, we further specialise to the case
$k_{1,3}=k_{1,-3}=k_{2,0}=k_{0,0}=k'$. Then dimensional reduction 
gives $K\cong  U(k')^4$,
\[U(8k')\longrightarrow U(k')^4, \]
and
$\phi^\pm_{n,m}$ are $k'\times k'$ complex matrices
acted on by some $SU(k')_L\times SU(k')_R$ subgroup.
The symmetry is further reduced by dynamical symmetry breaking
\[SU(8k')\longrightarrow SU(k')^4\times U(1)^3\longrightarrow SU(k')_{diag}\]
and the Higgs potential minimised by 
\[ \phi^\pm_{n,m}{}^0=\frac {\sqrt{3}}{2gr} U^\pm_{n,m},\]
where $U^+_{0,0}$, $U^+_{1,-3}$, $U^-_{2,0}$, $U^-_{1,-3}$ are four
unitary matrices satisfying one extra condition
\begin{equation} \label{HiggsCondition}
U^-_{2,0} U^+_{1,3} = U^+_{0,0} U^-_{1,-3}.
\end{equation}

\subsubsection{Quiver diagrams}

For a general $SU(3)$ irreducible representation, $\{l,\overline l\}$, 
the quiver diagram is
\begin{center}
\epsfxsize=6cm
\epsffile{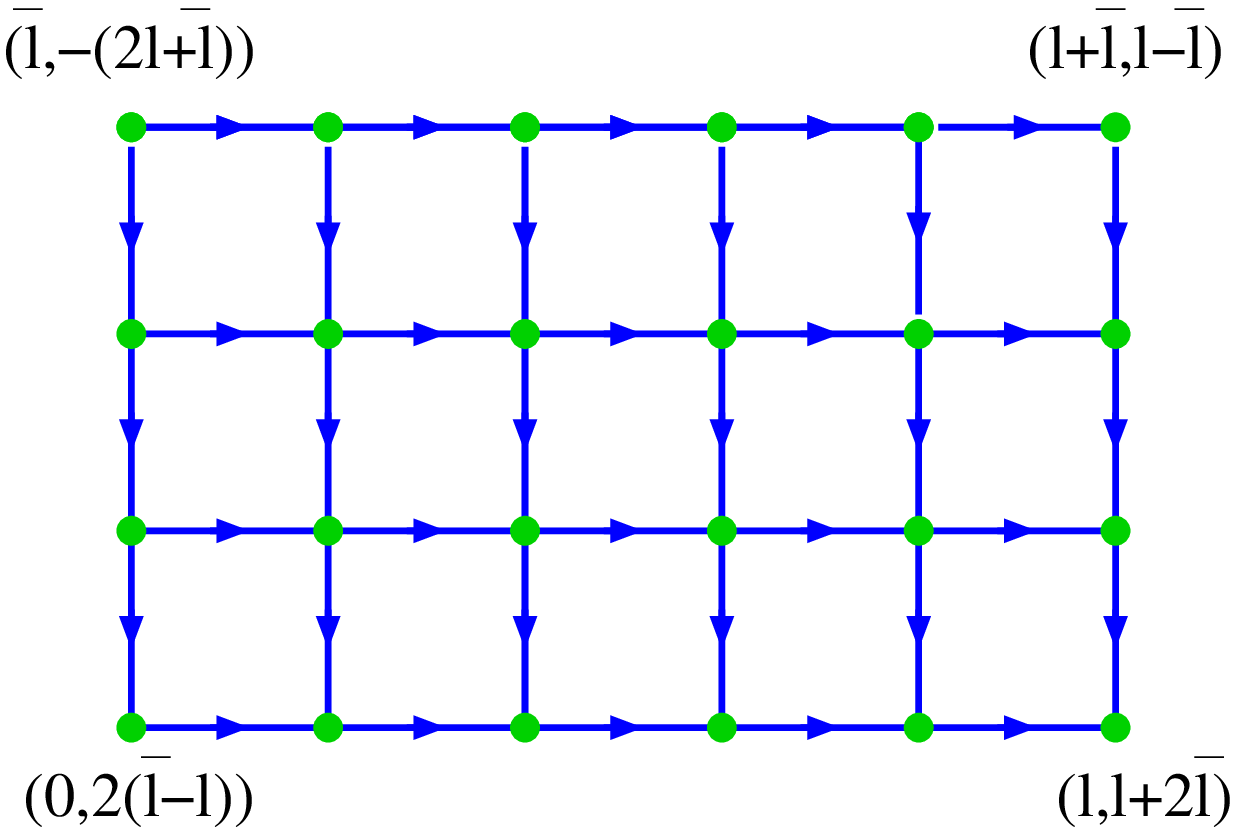}
\end{center}
The total number of Higgs matrices {(blue links)} is
$2l\overline l + l + \overline l$, while the
number of gauge groups {(green dots)} is $(l+1)(\overline l+1)$.
If $k_{n,m}=k'$ are all equal, then all Higgs fields are $k'\times k'$ matrices
and $V(\Phi)$ is minimised by those Higgs fields all proportional to
unitary matrices, with constraints of the form (\ref{HiggsCondition}) on the unitary matrices
around any plaquette.
Interpreting the Higgs fields as a $SU(k')$ lattice gauge field
on the quiver lattice, the constraints are satisfied by
demanding the 
trivial gauge configuration on the quiver lattice.

\section{Fermions and Yukawa couplings}

\subsection{Twisted Dirac operators on $\boldsymbol{S^2}$}

To study how dimensionally reduced fermions and Yukawa couplings
emerge in these models, we first consider the simplest non-trivial
example of $S^2$.
Represent the Dirac operator for a fermion with unit charge
in the presence of a magnetic monopole on $S^2$ of charge $p$ by 
$\Dirac^{(p)}_{S^2}$.
Mathematically, this is the Dirac operator twisted with the $p$-th tensor power of the
tautological line bundle $L$~\cite{Bott+Tu}.

For a given $p$, the eigenspinors will be denoted by $\chi_{j,p;l}$ and have
eigenvalues 
\[ \mu_{j,p}=\pm \frac 1 r \sqrt{\left(j+\frac {1+p} 2 \right)\left(j+\frac {1-p} 2 \right)}\]
so that
\[ \Dirac^{(p)}_{S^2} \chi_{j,p;l}(y)=\mu_{j,p} \chi_{j,p;l}(y).\]
For $p$ even the quantum number $j$ is half-integral 
while for odd $p$ it is integral: in both cases $j\ge {|p|+1\over 2}$
and the degeneracy is $2j+1$, labelled by $l=0,1,\ldots,2j$.
The eigenspinors can be decomposed into their positive and
negative chirality components 
\[\chi_{j,p;l}=\begin{pmatrix}\chi^+_{j,p;l}\\ \pm \chi^-_{j,p;l} \end{pmatrix},\]
where the sign corresponds to the sign of the eigenvalue.

In addition, for the special value $j=\frac{|p|-1} 2$ 
when $p\ne 0$,
there are $|p|$ zero modes:
for $p\ge 1$ there are $p$ negative chirality modes, 
which we denote by
\[ \chi^-_{p;r}, \qquad r=0,1,\ldots,p-1,\]
while for  $p\le -1$ there are $|p|$ positive chirality modes, 
\[ \chi^+_{p;r}, \qquad r=0,1,\ldots,|p|-1.\]
For a given monopole charge, the index of the Dirac operator is
\[\hbox{Index}\bigl(\Dirac_{S^2}^{(p)}\bigr)=-p. \]

The Dirac operator on $\M_6$ splits up into the direct sum of
4-dimensional and 2-dimensional Dirac operators 
\[\Dirac_{(6)}=\Dirac_{(4)}\otimes {\bf 1}_2 + \gamma_5\otimes \Dirac_{S^2}.\]
At first sight zero modes of the Dirac operator on $S^2$ 
might be expected to manifest themselves as massless 
fermions for the Dirac operator on $\M_4$, but we shall see below that this is
not always the case.

After dimensional reduction a fermion on $\M_6$,
e.g. in the fundamental of $U(k)$, will decompose as
\[ \Psi(x,y)=\begin{pmatrix} \Psi^+(x,y) \\ \Psi^-(x,y) \\ \end{pmatrix}\]
where the $\pm$ signs refer to the $S^2$ chirality, not 4-dimensional or 6-dimensional chirality.
Indeed $\Psi$ itself could be either Dirac or Weyl in 6-dimensions.
In the equivariant dimensional reduction framework only zero modes on $S^2$
are compatible with $SU(2)$ symmetry: $j>{|p|-1\over 2}$ correspond to
higher harmonics which do not have this symmetry and correspond to
4-dimensional fermions with masses of order ${1\over r}$.  Focusing on zero modes, the
6-dimensional fermions $\Psi^\mp$ decompose as
\begin{eqnarray}\Psi^-(x,y)&=& \oplus_{r=0}^{p_i-1} 
\widetilde\psi_{p_i;r}(x)\chi^-_{p_i;r}(y),\ \Psi^+=0\quad 
{(p_i\ge 1}) , \nonumber \\
\Psi^+(x,y)&=&\oplus_{r=0}^{|p_i|-1} 
\psi_{p_i;r}(x)\chi^+_{p_i;r}(y),\ \Psi^-=0\quad {(p_i\le -1)},\nonumber \end{eqnarray}
where 
$\widetilde\psi_{p_i;r}(x)$  and $\psi_{p_i;r}(x)$ are either Dirac spinors
in 4-dimensions, if $\Psi$ is Dirac in 6-dimensions, or
Weyl spinors of opposite chirality, if $\Psi$ is Weyl in 6-dimensions.

Not all of the 4-dimensional fermions $\widetilde\psi_{p_i;r}(x)$
and $\psi_{p_i;r}(x)$ are massless however~\cite{DR1}.
In 6 dimensions the Dirac operator involves the 6-dimensional gauge field, which
includes the Higgs field after dimensional reduction, and
these induce 4-dimensional Yukawa couplings, allowing for the possibility of 
generating
mass terms for 4-dimensional fermions through dynamical symmetry breaking. 
If, and only if,  $m$ is odd there is a
$4$-dimensional Yukawa coupling linking $\widetilde\psi_{1}$ to $\psi_{-1}$ through
\[ \frac g 2
\int_{\M_4}\sqrt{-g_{(4)}}\,\phi^\dagger_{\frac{m+1}2}
\overline{\psi_{-1}} \,\gamma_5\, \widetilde\psi_{1}\, d^4x +h.c.\]
For the example in \S\ref{threetwoone},  $SU(3)\rightarrow SU(2)\times U(1)\rightarrow U(1)$, we had
$k_0=2$, $k_1=1$, and $m=1$.
In this case
$\widetilde\psi_{1}$ transforms as ${\bf 2}_1$ under $SU(2)\times U(1)$,
$\psi_{-1}$ as ${\bf 1}_{-2}$, and
$\phi=\phi_1$ as ${\bf 2}_1$. 
These 4-dimensional fermions pick up a mass $\frac 1 {2r}$ via the Higgs vacuum expectation value,
which is of the same order as the masses of the higher harmonic fermions
arising from non-zero eigenvalues of the Dirac operator on $S^2$
and therefore should be removed from consideration if we are assuming higher
harmonics are too heavy to be relevant to the physics at low energies.

\subsection{Spin$^{\boldsymbol c}$ structures on $\boldsymbol{\CP2}$}

The issue of fermions on $\CP2$ is complicated because there is
a topological obstruction to the existence of a spin structure:
due to the fact that the second Stiefel-Whitney class is non-vanishing~\cite{Milnor} there is a global obstruction to defining spinors on ${\bf CP}^n$
for even $n$.

Nevertheless, fermions can be defined by coupling them to monopoles and/or 
instantons (spin${}^c$ structures). 
The full spectrum of the twisted Dirac operator is complicated
but for equivariant dimensional reduction we only need the zero modes.
For fermions coupling to an equivariant monopole of magnetic charge $m$ and an 
equivariant instanton
of topological charge $n$, the index of the Dirac operator on $\CP2$ is~\cite{DR2}
\[\hbox{Index}\big(\Dirac^{(n,m)}\big)=\frac 1 8 (n+1) \left(m^2-(n+1)^2\right) .
 \]
The fact that this is not an integer if $n$ and $m$ have the same parity,
{i.e.} they are either both even or both odd (e.g. $n=m=0$), is related to the lack of spin structure on $\CP2$.
Under the embedding $SU(2)\times U(1)\hookrightarrow SU(3)$,
$\{l,\overline l\}\rightarrow \oplus(n,m)=:W_{l,\overline l}$, $n$ and
$m$ always have the same parity, so any equivariant monopole/instanton
background arising from the embedding will not admit global spinors.
We therefore allow for a further twist with a monopole of charge $q\in {\bf Z}+\frac 1 2$ ($2q$ odd) and the index for this twisted gauge field configuration is
\[\hbox{Index}\big(\Dirac_q^{(n,m)}\big)=\frac 1 8 (n+1) \left((m+2q)^2-(n+1)^2\right).
 \]
We shall denote the positive and negative chirality zero modes of this
operator, with a given fixed $q$, by 
$\chi^+_{n,m,q}$ and $\chi^-_{n,m,q}$ respectively (for 
notational clarity the degeneracy is not indicated).

\subsubsection{Fundamental representation}

For $\{l,\overline l\}=\{1,0\}$ we have
$\{1,0\}\rightarrow (1,1)\oplus (0,-2)$, and choosing for example
$q=-\frac 1 2$ results in
\[\hbox{Index}\big(\Dirac_{-1/2}^{(1,1)}\big)=-1,\hskip 20pt  
\hbox{Index}\big(\Dirac_{-1/2}^{(0,-2)}\big)=1. \]
For example, the case
$k=3k'$ with $k_{1,1}=k_{0,-2}=k'$ gives a 
single $k'\times k'$ Higgs matrix and the symmetry reduction scheme
\[ SU(3k')\longrightarrow SU(k')\times SU(k')\times U(1)
\longrightarrow SU(k').\]
With
$2q=-1$, $\chi^+_{0,-2,-\frac 1 2 }(y)$ and $\chi^-_{1,1,-\frac 1 2}(y)$ 
are the only zero modes giving the equivariant decomposition
\[ 
\Psi=\begin{pmatrix} \psi_{0,-2}(x)\chi^+_{0,-2,-\frac 1 2}(y) \\ 
\widetilde\psi_{1,1}(x)\chi{}_{1,1,-\frac 1 2}^-(y) \\ \end{pmatrix},
\] 
where $\psi_{0,-2}(x)$ and $\widetilde\psi_{1,1}(x)$ are either 4-dimensional
Dirac spinors on $\M_4$, if $\Psi$ is Dirac in 8-dimensions,
or chiral spinors
of opposite chirality in 4-dimensions, if $\Psi$ is chiral in 8-dimensions.
The induced 4-dimensional Yukawa couplings generate a mass term
for these spinors given by
\[ \frac {\sqrt{2}}  {r} \big(\psi_{0,-2}^\dagger \gamma_5\,\widetilde\psi_{1,1}
+ \widetilde\psi_{1,1}^\dagger \gamma_5 \,\psi_{0,-2}\big).
\] 

A different choice of $q$ leads to a different conclusion.
Taking $2q=3$ results in
\[\hbox{Index}\big(\Dirac_{3/2}^{(1,1)}\big)=3,\hskip 20pt  
\hbox{Index}\big(\Dirac_{3/2}^{(0,2)}\big)=0.\]
There is no analogue of $\psi_{0,-2}(x)$ in this case
and Yukawa couplings cannot generate a mass term in 4-dimensions.

\subsubsection{Adjoint representation}

Starting from the adjoint representation
\[
\{l,\bar l \}=\{1,\overline 1\}\longrightarrow (2,0)\oplus(1,3)\oplus(1,-3)\oplus(0,0) ,
\]
consider the symmetry breaking scheme
\[ SU(8k')\longrightarrow SU(k')^4\times U(1)^3\longrightarrow SU(k').\]
Choosing, for example, $q=-\frac 3 2$ gives
\begin{eqnarray}
\hbox{Index}\big(\Dirac_{-3/2}^{(2,0)}\big)&=&0,\hskip 20pt  
\hbox{Index}\big(\Dirac_{-3/2}^{(1,3)}\big) \ = \ -1,\nonumber \\  
\hbox{Index}\big(\Dirac_{-3/2}^{(1,-3)}\big)&=&8,\hskip 20pt  
\hbox{Index}\big(\Dirac_{-3/2}^{(0,0)}\big) \ = \ 1. \nonumber
\end{eqnarray}
In this case Yukawa couplings generate a mass coupling the 4-dimensional spinors
$\widetilde\psi_{1,3}(x)$ and $\psi_{0,0}(x)$, 
but the {8 flavours} $\psi_{1,-3}(x)$ remain massless.

\section{Conclusions}

We have shown that equivariant dimensional reduction with a simple gauge group $G$
gives the following:
\begin{itemize}
\item Gauge symmetry reduction
$G\rightarrow K$ with only one gauge coupling in 4-dimensions, 
even if $K$ is semi-simple.
\item  Further dynamical symmetry breaking $K\rightarrow K'$
where the vacuum and symmetry breaking patterns, including
Higgs and gauge boson masses and Weinberg angles, can be deduced uniquely
from group theory and induced representation theory.
\item  In certain cases the vacuum configuration is related to gauge dynamics on the quiver
lattice: the Higgs vacuum corresponds to zero flux on the quiver lattice.
\item When fermions are included, 
chiral theories with families emerge naturally from 
non-trivial fluxes on~$S/R$.
\item Chiral fermions on $\M_d$ do not allow direct mass terms, but Yukawa
couplings can give 4-dimensional masses to some of the resulting fermions on $\M_4$.  
Yukawa couplings can even give masses to some, but not all, zero modes.
\end{itemize}
The gauge and fermion structure of equivariant dimensionally
reduced field theories is clearly very rich.
Standard model type Yukawa couplings, with different chiralities belonging
to different irreducible representations of the gauge group, arise
quite naturally in the models presented here, but an exhaustive
analysis of all possibilities would be an ambitious programme
and remains to be tackled.

\subsection*{Acknowledgments}

We thank A.~Chatzistavrakidis and H.~Steinacker for helpful discussions. The work of BPD is supported in part by the EU 
Research Training Network in Noncommutative Geometry (EU-NCG). The work of RJS is supported in part by grant ST/G000514/1 ``String Theory
Scotland'' from the UK Science and Technology Facilities Council.


\begin{thebibliography}{99}
%\addtolength{\itemsep}{-2pt}

\bibitem{KK} Th.~Kaluza, Sitzungsber. Preuss. Akad. Wiss. K {\bf 1}
  (1921) 966; 
O.~Klein, Z. Phys. {\bf 37} (1926) 895.
% Kaluza, Theodor (1921). "Zum Unittsproblem in der
% Physik". Sitzungsber. Preuss. Akad. Wiss. Berlin. (Math. Phys.)
% 1921: 966972.  
% Klein, Oskar (1926). "Quantentheorie und fnfdimensionale
% Relativittstheorie". Zeitschrift fr Physik a Hadrons and Nuclei 37
% (12): 895906 

\bibitem{KKReview}
T.~Appelquist, A.~Chodos and P.G.O.~Freund, {\sl Modern Kaluza-Klein
  Theories} (Addison-Wesley, 1987).

\bibitem{FM}
P.~Forg\'acs and N.S.~Manton, Commun. Math. Phys. {\bf 72} (1980) 15;
C.H.~Taubes, Commun. Math. Phys. {\bf 75} (1980) 207.

\bibitem{ThCh}
A.~Chatzistavrakidis, these proceedings.

%\cite{ShelterIsland}
\bibitem{ShelterIsland}
E.~Witten, in: {\sl Proceedings of the 1983 Shelter Island Conference
on Quantum Field Theory and the Fundamental Problems of Physics},
eds. R.~Jackiw, N.N.~Khuri, S.~Weinberg and E.~Witten (MIT Press,
1985), p.~227.

\bibitem{George}
  D.~Kapetanakis and G.~Zoupanos,
  %``Coset Space Dimensional Reduction Of Gauge Theories,''
  Phys.\ Rept.\  {\bf 219} (1992) 1.
  %%CITATION = PRPLC,219,1;%%

\bibitem{Yuri} Y.A.~Kubyshin, J.M.~Mourao, G.~Rudolph and I.P.~Volobujev, 
{\sl Dimensional Reduction of Gauge Theories, Spontaneous
  Compactification and Model Building} (Springer, 1989).

%\cite{ALCP}
\bibitem{ALCP}
L.~\'{A}lvarez-C\'onsul and O.~Garc\'{\i}a-Prada,
%``Dimensional reduction and quiver bundles,''
J. Reine Angew. Math. {\bf 556} (2003) 1
[arXiv:math.DG/0112160];
%``Hitchin-Kobayashi correspondence, quivers and vortices,''
Commun. Math. Phys. {\bf 238} (2003) 1 
[arXiv:math.DG/0112161].
%%CITATION = MATH.DG 0112161;%%
%%CITATION = MATH.DG 0112160;%%

\bibitem{LPS}
  O.~Lechtenfeld, A.D.~Popov and R.J.~Szabo,
%  ``Quiver gauge theory and noncommutative vortices,''
  Progr. Theor. Phys. Suppl. {\bf 171} (2007) 258
  [arXiv:0706.0979~[hep-th]].
  %%CITATION = ARXIV:0706.0979;%%

%\cite{DR1}
\bibitem{DR1}
  B.P.~Dolan and R.J.~Szabo,
  %``Dimensional Reduction, Monopoles and Dynamical Symmetry
  %Breaking,''
  JHEP {\bf 03} (2009) 059
  [arXiv:0901.2491 [hep-th]].
  %%CITATION = ARXIV:0901.2491;%%

\bibitem{DR2}  B.P.~Dolan and R.J.~Szabo,
% Dimensional Reduction and Vacuum Structure of Quiver Gauge Theory
  JHEP {\bf 08} (2009) 038 [arXiv:0905.4899 [hep-th]].
  %%CITATION = ARXIV:0905.4899;%%

\bibitem{Bott+Tu}
  R.~Bott and L.W.~Tu,
  {\sl Differential Forms in Algebraic Topology} (Springer, 1982).

\bibitem{Milnor}
J.W. Milnor and J.D. Stasheff, {\sl Characteristic Classes} (Princeton
University Press, 1974).

\end{thebibliography}
\end{document}